\documentclass[twocolumn,prx,superscriptaddress,floatfix,longbibliography]{revtex4-2}
\usepackage{graphicx}
\usepackage{xcolor}
\usepackage{hyperref}
\usepackage{braket}
\usepackage{bm}
\usepackage{bbold}
\usepackage{soul}

\expandafter\let\csname equation*\endcsname\relax 
\expandafter\let\csname endequation*\endcsname\relax
\usepackage{amsmath}
\usepackage{cleveref}

\usepackage{ulem}  
\begin{document}
	
	\title{Statistical learning on randomized data to verify quantum state approximate $k$-designs}
%

	\author{Kaustav Mukherjee}
	\affiliation{Department of Physics and Astronomy, University of Tennessee, Chattanooga, TN 37403, USA}
	\affiliation{UTC Quantum Center, University of Tennessee, Chattanooga, TN 37403, USA}	\email{Kaustav-Mukherjee@utc.edu}
	
	\author{Sarah Chehade}
		\affiliation{Department of Physics and Astronomy, University of Tennessee, Chattanooga, TN 37403, USA}
		\affiliation{UTC Quantum Center, University of Tennessee, Chattanooga, TN 37403, USA}
		
		\author{Lorenzo Versini}
		\affiliation{Blackett Laboratory, Imperial College London, SW7 2AZ, London, UK}
		\affiliation{Atomic and Laser Physics (ALP), Department of Physics, University of Oxford, Parks Road, Oxford OX1 3PU, United Kingdom}
		
				\author{Karim K. Alaa El-Din}
		\affiliation{Blackett Laboratory, Imperial College London, SW7 2AZ, London, UK}
		\affiliation{Atomic and Laser Physics (ALP), Department of Physics, University of Oxford, Parks Road, Oxford OX1 3PU, United Kingdom}
		
						\author{Florian Mintert}
		\affiliation{Blackett Laboratory, Imperial College London, SW7 2AZ, London, UK}
		\affiliation{Helmholtz-Zentrum Dresden-Rossendorf, Bautzner Landstra{\ss}e 400, 01328 Dresden, Germany}
		
			\author{Rick Mukherjee}
					\affiliation{Department of Physics and Astronomy, University of Tennessee, Chattanooga, TN 37403, USA}
					\affiliation{UTC Quantum Center, University of Tennessee, Chattanooga, TN 37403, USA}
							\affiliation{Blackett Laboratory, Imperial College London, SW7 2AZ, London, UK}

	\vspace{10pt}

	\begin{abstract}
		Random ensembles of pure states have proven to be extremely important in various aspects of quantum physics such as benchmarking the performance of quantum circuits, testing for quantum advantage, studying many-body thermalization and black hole information paradox. Although generating a truly random quantum ensemble is experimentally challenging, approximate realizations are equally valuable and are known to emerge naturally in a variety of physical models, including Rydberg setups. These are referred to as approximate quantum state designs, and verifying their degree of randomness can be a measurement intensive task, similar to performing full quantum state tomography on many-body systems. In this theoretical work, we present a measurement scheme and analysis techniques to validate the degree of randomness of a quantum ensemble generated by a simulated experimental setup. This is achieved by translating the information residing in the complex many-body state into a succinct representation of classical data using projective measurements in randomly chosen bases, which is then processed using methods of statistical inference such as maximum likelihood estimation and neural networks, benchmarked against the predictions of shadow tomography. Our scheme only requires individually addressed single qubit operations to be performed in order to be employed, making it applicable for a range of physical platforms.
	\end{abstract}
	
	\maketitle
	
	\section{Introduction}
	Random number generation has a wide range of applications, including information security \cite{Schneier}, networking \cite{Guang, Seetharam}, board games, lotteries, and gambling \cite{hull1962random, marsaglia2003random}. However, it is very costly and almost impossible to generate a genuinely random string of numbers, in the deterministic classical world. For practical purposes, pseudorandom number generators are used where a deterministic computer algorithm creates a sequence of numbers that appear to be statistically random when compared to a fully independent set of random variables \cite{Marsaglia1990, James1990, park1988random, Blum1986}. Thus, the average value of the pseudorandom distribution will not be distinguishable from the average value of a perfect random distribution. Similarly, in combinatorial mathematics, one can define classical $k$-designs which are indistinguishable from truly random distributions up to $k$-moments.  \cite{bannai1979tight, bannai2009survey,emerson2003pseudo, stinson2004combinatorial}.
	
		\begin{figure}[htb]
		\centering
		\includegraphics[width=0.99\linewidth]{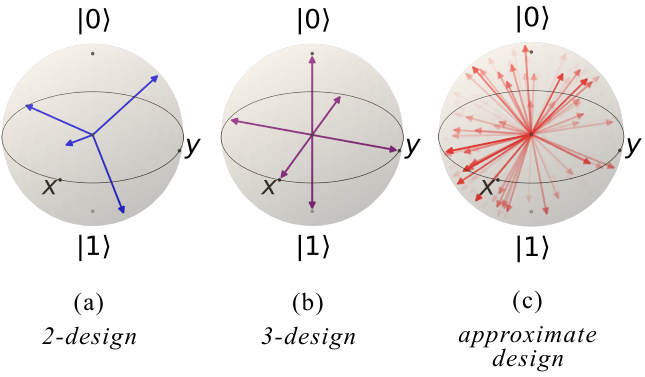}
		\caption{The figure depicts the quantum k-designs for a single qubit in a Bloch sphere for $k=2,3$ in (a),(b) and an ensemble of states that approximates an arbitrary k-design in (c). Less transparent vectors in (c) occur with higher probability within the ensemble of states.}
		\label{fig:kdes-bloch}
	\end{figure}
	
	Quantum $k$-designs are extensions of the classical $k$-design concept in which the probability distributions of unitary operators or quantum states averaged over a polynomial function up to degree $k$ is identical to the average over a unique group-invariant measure, in this case a Haar measure. Quantum $k$-designs find many applications in quantum information theory which include estimating fidelity \cite{Dankert2009}, performing state tomography \cite{Ohliger2013}, implementing cryptography \cite{Adam2013, Lancien2020}, randomized encoding and benchmarking of quantum circuits \cite{cross2019validating, brandao2021models, boixo2018characterizing, bouland2019complexity, haferkamp2020closing} and simulators \cite{bernien2017probing}, checking for entanglement \cite{brydges2019probing}, generating topological order \cite{weinstein2008parameters} and studying the black hole information paradox \cite{hayden2007black, piroli2020random}. Recent studies also reveal the relevance of quantum $k$-designs in the thermalization process for many-body systems \cite{Choi2021, Cotler2021, nautiyal2023quantum}. However, generating quantum designs on any quantum processor is challenging and requires a high degree of control over a large set of noisy parameters \cite{Dankert2009, brandao2016local, Ohliger2013, onorati2017mixing, nakata2017efficient, elben2018renyi}, as the number of gates grows exponentially with the number of qubits. Alternatively, \textit{approximate quantum state designs} (see Fig.~\ref{fig:kdes-bloch}) are relatively easy to construct and by definition, they are distributions of states that approximate their exact counterparts.

	The degree of randomness in an ensemble of states is characterized by estimating its closeness to the relevant quantum $k$-design, a task that naturally depends on the information available about the ensemble of states. However, gathering such information can be an experimentally resource intensive task, akin to quantum state tomography for large systems \cite{White, Lvovsky, Itatani, Cramer}. Over the years, there has been a push towards efficient quantum state tomography \cite{Lanyon2017, Cramer2010, huang2020predicting, Smith2021}. In the same spirit, this work deals with the efficient and accurate characterization of numerically simulated random ensembles of states.
	
	The characterization of approximate quantum $k$-designs investigated here involves two key steps: (i) Sampling the ensemble of states using projective measurements in the random bases and (ii) applying statistical learning methods on the measurement data for post-processing. Using a prototypical platform of interacting spins in a Rydberg setup, we numerically construct the ensemble of states by implementing a protocol that involves splitting the system into two subsystems: $A$ and $B$, realizing an approximate quantum state design in $A$ and treating $B$ as an auxiliary system used for random ensemble generation~\cite{Choi2021}. Measuring subsystem $B$ in the computational basis projects the joint system into a corresponding ensemble of post-measurement states on subsystem $A$. We then propose randomly selecting the basis of projective measurement for subsystem $A$ while measuring subsystem $B$ in the computational basis. It is well-established that utilizing the statistical correlations arising from randomized measurements forms a flexible toolbox useful for estimating various quantum properties of the many-body quantum state such as purity, entanglement and out-of-time-ordered correlations \cite{Elben}. Next, we leverage tools of machine learning (ML) which are based on statistical inferences to estimate the parameters that describe the ensemble of states. Using both maximum likelihood estimation and restricted Boltzmann machines, we demonstrate that this routine significantly reduces the number of measurements needed to accurately represent the ensemble of states compared to either full state tomography or a frequentist method as implemented in~\cite{Choi2021}. In addition, we show that the maximum likelihood approach outperforms even efficient shadow tomography~\cite{o2016efficient,huang2020predicting, McGinley} for all cases tested, while the restricted Boltzmann machines achieve this performance in the high data regime. Finally, we note that our results are applicable to any quantum platform that can generate random ensembles of states.
	
	\section{Theory}
	\subsection{Constructing approximate quantum state $k$-design using projected ensembles}
	A uniform distribution of all pure states in a Hilbert space is defined by the Haar ensemble \cite{Adam2013}. Averaging arbitrary functions across the ensemble is impossible in practice, due to the finite number of samples that can be collected. Instead, it is possible to use quantum state designs that reproduce the universal properties of randomness of the Haar ensemble up to a certain order.  However, these designs are typically constructed approximately using random unitary quantum circuits \cite{schuster2025random,brandao2016local, harrow2009random, nakata2017efficient, riddell2025quantum}, which requires sufficient depth of the quantum circuit, resulting in polynomial scaling with the number of qubits, thereby contributing to the overall infidelity of the operation.
	
	Another approach to generating approximate quantum state designs involves making local measurements on many-body states that evolve under a time-dependent Hamiltonian in quantum many-body systems \cite{Cotler2021, Choi2021}. This approach of constructing approximate $k$-designs, which is also referred to as the \textit{projected ensemble method}, is advantageous due to its generality, as it can be readily implemented in experimental platforms that realize Ising models  \cite{Choi2021}. We consider a system of $N$ spins, which are partitioned into two subsystems: $A$ and $B$, with $N_A$ and $N_B$ spins, respectively such that $N_A \ll N_B$. In this work, we numerically simulate the works  \cite{Cotler2021, Choi2021} to generate the data for approximate quantum state $k$-designs, which is then characterized using statistical learning methods.
	
	 Using the projective measurement approach, an ensemble of pure states is generated by performing local measurements on subsystem $B$ for given projected states in complementary subsystem $A$. The naive brute-force approach to characterize the ensemble of states would be to  perform measurements in all directions ($X,Y,Z$), on both  subsystems $A$ and $B$. This allows for reconstruction of the full many-body state, which  captures the complete information about the entire system. However, this procedure is prohibitively expensive, especially for larger systems. Alternatively, one may measure both the subsystems, solely in the computational or $Z$ basis as  performed in the original experiment \cite{Choi2021}. Measuring the entire system in the computational basis results in the loss of useful information (such as phase), which  can be useful for characterizing the ensemble of pure states.  Instead, we propose measuring each spin in subsystem $A$ in a randomly chosen basis from $X$, $Y$, or $Z$, while measuring subsystem $B$ uniformly in the $Z$ basis. This measurement scheme results in bitstrings $r_A\in\{r_{A}^{(1)},r_{A}^{(2)},...\}$ for $r_A^{(i)}\in\{x_A^{(i)},y_A^{(i)},z_A^{(i)}\}$ for subsystem $A$ and $z_{B}$ for subsystem $B$. Each bitstring contains a string of bits 0 or 1 representing the state of the individual spin when measured in the chosen basis. The scheme of measurements used in A is referred to as  \textit{measurement in a randomly chosen basis} as  discussed in more detail in Appendix B. The many-body state of the full system is denoted by $|\Psi\rangle$ and the state of subsystem $A$ conditioned that a particular bitstring $z_B$ has been measured is given by 
	\begin{equation}
		|\Psi_A(z_B)\rangle = (\mathbb{1}_A \otimes \langle z_B|)|\Psi\rangle/\sqrt{p(z_B)} ,
	\end{equation}
	where $p(z_B) = \langle \Psi| (\mathbb{1}_A \otimes |z_B\rangle \langle z_B|)|\Psi\rangle$ is the probability to measure a specific bitstring $z_B$. The marginal probability of measuring the subsystem $A$ in a particular state $r_A$ for a given choice of measurement basis is given by
	\begin{align}\label{marginal}
		p(r_A) &= \sum_{z_B} p(z_B) p(r_A|z_B) ,\nonumber \\
		&= \sum_{z_B} p(z_B)  |\langle r_A|\Psi_A(z_B)\rangle|^2,
	\end{align}
	where $p(r_A|z_B)$ is the conditional probability distribution. Thus, the ensemble of states is a collection of all such conditional states of subsystem $A$ along with its corresponding probabilities to measure a given bitstring $z_B$ and is written as
	\begin{equation}\label{eos}
		\mathcal{E} = \{(|\Psi_A(z_B)\rangle, p(z_B))\}\qquad \forall\ z_B \in [0, 2^{N_B}) .
	\end{equation}
	The $k^{\text{th}}$ moment of the ensemble $\mathcal{E}$ is defined as follows
	\begin{equation}\label{eq:moment-ens}
		\rho^{(k)}_{\mathcal{E}} = \sum_{z_B} p(z_B)\ (|\Psi_A(z_B)\rangle\langle \Psi_A(z_B)|)^{\otimes k} ,
	\end{equation}
	while the $k^{\text{th}}$ moment of the Haar ensemble for a Hilbert space of dimensionality $d$ is defined as  $ \rho^{(k)}_{Haar} = \int_{\psi\sim \rm{Haar}(d)} d\psi (|\psi\rangle\langle \psi|)^{\otimes k}$, which has a closed analytical form as detailed in Appendix \ref{AnaHaar}. Here, $\psi$ refers to random pure states that are obtained from the unitarily invariant (Haar) measure on the unit sphere in $d$-dimensional Hilbert space. The degree of closeness between $\mathcal{E}$ and the Haar ensemble is quantified using the 1-norm i.e., the trace distance, between the two ensembles which is defined as
	\begin{equation}\label{eq:tr-distance}
		\delta_{(k)} = \frac{1}{2}\mathrm{Tr}\left(\sqrt{\left(\rho_{\mathcal{E}}^{(k)} - \rho_{Haar}^{(k)}\right)^{\dagger}\left(\rho_{\mathcal{E}}^{(k)} - \rho_{Haar}^{(k)}\right)}\right) .
	\end{equation}
	Thus, for $\delta_{(k)} \ll 1$, the many-body state $|\Psi\rangle$ yields an ensemble of pure states on the subsystem whose probability distribution is indistinguishable from the Haar ensemble up to $k$-th moment, thereby forming an approximate quantum state $k$-design. The efficient and accurate characterization of (approximate) quantum state $k$-designs which is related to estimating the value of $\delta_{(k)}$ depends on the type of information encoded in the ensemble of states $\mathcal{E}$. This in turn relies on the actual details of the measurement protocol. 
	
	In this work, we use four different protocols to characterize quantum designs: (i) frequentist method (ii) shadow tomography \cite{Huang} (iii) complex-restricted Boltzmann machine (cRBM) \cite{Carleo2017,park2022expressive} and (iv) maximum likelihood estimation (MLE) of which the latter two demonstrate greater efficiency with respect to data acquisition. Each of the four protocols uses a distinct ansatz to reconstruct the state $|\Psi_A(z_B)\rangle$ for each $z_B$ from repeated measurement outcomes: (i) the frequentist method reconstructs the state by determining the coefficients of $|\Psi(z_B)\rangle$ based on probabilities of outcomes in the computational basis, (ii) shadow tomography estimates the density operator from measurement outcomes in random bases leveraging unbiased statistics (U-statistics, see Appendix \ref{shadowtom}), (iii) the complex restricted Boltzmann machine (cRBM) employs a bi-partite neural network to parameterize the state from measurements in a randomly chosen basis , and, (iv) the maximum likelihood estimation (MLE) directly optimizes the coefficients of $|\Psi(z_B)\rangle$ from measurement outcomes in random basis. Table~\ref{table_1} summarizes the measurement basis, state ansatz and estimation errors, highlights how the error decreases with the size of the measurement dataset for each protocol. The underlying theory of these four protocols and their implementation for this work are described in detail in \Cref{estimators}. In the next section, we describe the theory behind Rydberg simulator used to generate the approximate k-designs similar to \cite{Cotler2021, Choi2021}.

	\subsection{Generating random ensemble states with Rydberg simulator}
	Rydberg atoms are neutral atoms with their valence electron placed in a highly excited state, possessing large dipole moments \cite{gallagher2006rydberg,saffman2010quantum}. The characteristic strong dipole-dipole interaction between neighboring Rydberg atoms has been exploited for a wide variety of applications that include many-body physics \cite{Adams2020,browaeys2020many,bluvstein2021controlling}, quantum simulation \cite{Browaeys2016,mukherjee2020two,scholl2021quantum} and quantum computing\cite{Saffman2016,henriet2020quantum}. In this work, a linear chain of trapped atoms is considered, which are initialized to a state with all atoms in their ground state $|g\rangle$. The laser with detuning  $\Delta$ and Rabi frequency $\Omega$ excites the individual atoms to their Rydberg state, denoted by $|e\rangle$. The system evolves according to the time-independent Hamiltonian
	\begin{equation}\label{eq:hamiltonian}
		\hat H = \frac{\hbar\Omega}{2}\sum^{N}_{i=1} \hat{\sigma}_i^x - \hbar\Delta\sum^{N}_{i=1} \hat{n}_i + \frac{C_6}{a^6} \sum_{j>i}\frac{\hat{n}_i \hat{n}_j}{|i-j|^6},
	\end{equation}
	where $\hat{\sigma}_i^x =  |g\rangle_i\langle e |_i +  |e\rangle_i\langle g |_i$ and $\hat{n}_i= |e\rangle_i\langle e |_i$. Two atoms in their Rydberg states interact with each other via the van der Waals interaction, whose strength is given by the van der Waals coefficient $C_6$. Simultaneous excitations of nearest neighboring atoms to their Rydberg states are suppressed due to the Rydberg blockade mechanism \cite{urban2009observation}. This effect allows us to work with the reduced Hilbert space, which makes the numerical simulations more tractable. 
	
	\begin{figure}[t]
		\includegraphics[width=\linewidth]{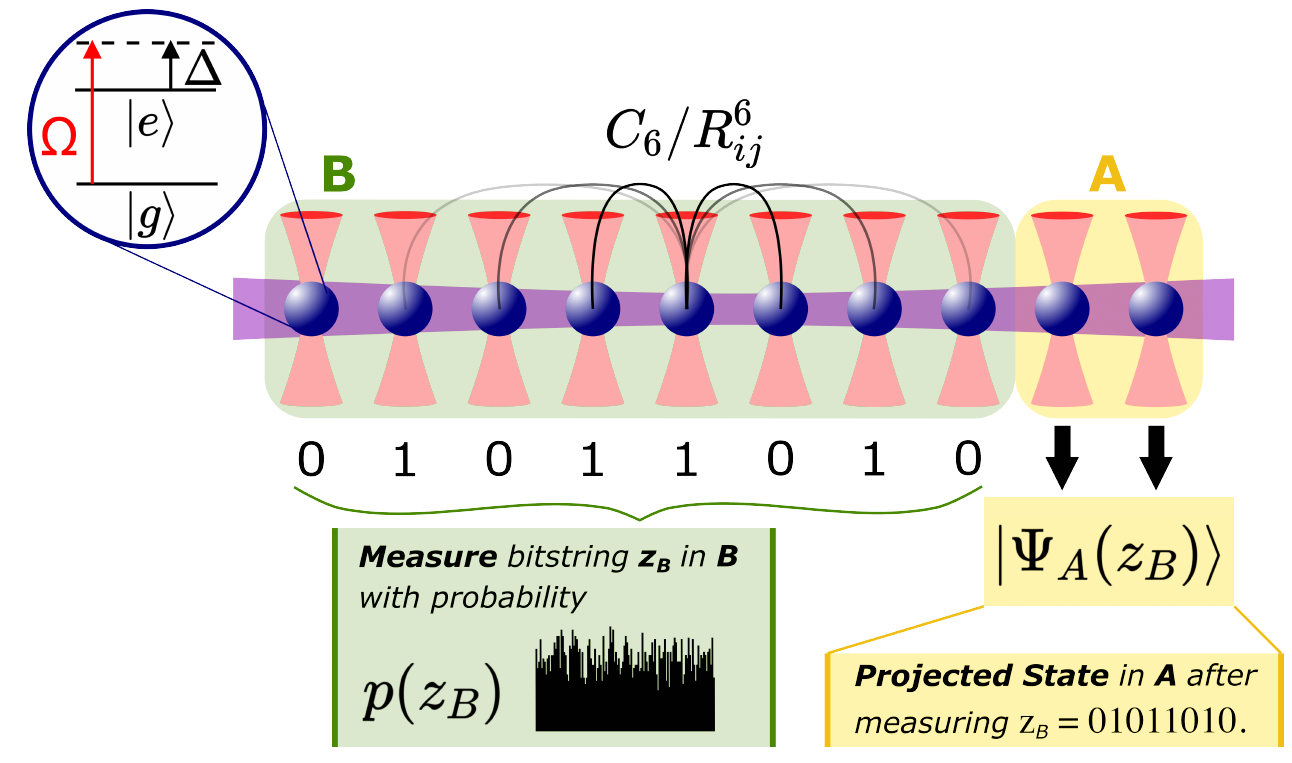}
		\caption{Simulated Rydberg setup: $N=10$ Rydberg atoms (blue spheres) trapped by optical tweezers (red cones) which are globally driven by a laser beam (purple) characterized by Rabi frequency $\Omega$ and detuning $\Delta$ (shown as inset). The first 8 atoms are treated as subsystem $B$ whose projective measurements in the computational basis provide the bitstring $z_B$. The remaining two atoms form the subsystem $A$ described by the wave function $|\Psi_A(z_B)\rangle$ which depends on the outcome $z_B$.}
		\label{fig:system-diagram} 
	\end{figure}
	
	\subsection{Numerical simulation of the Rydberg experiment \cite{Choi2021}}
	
\begin{figure}[t]
		\centering
		\includegraphics[width=0.99\linewidth]{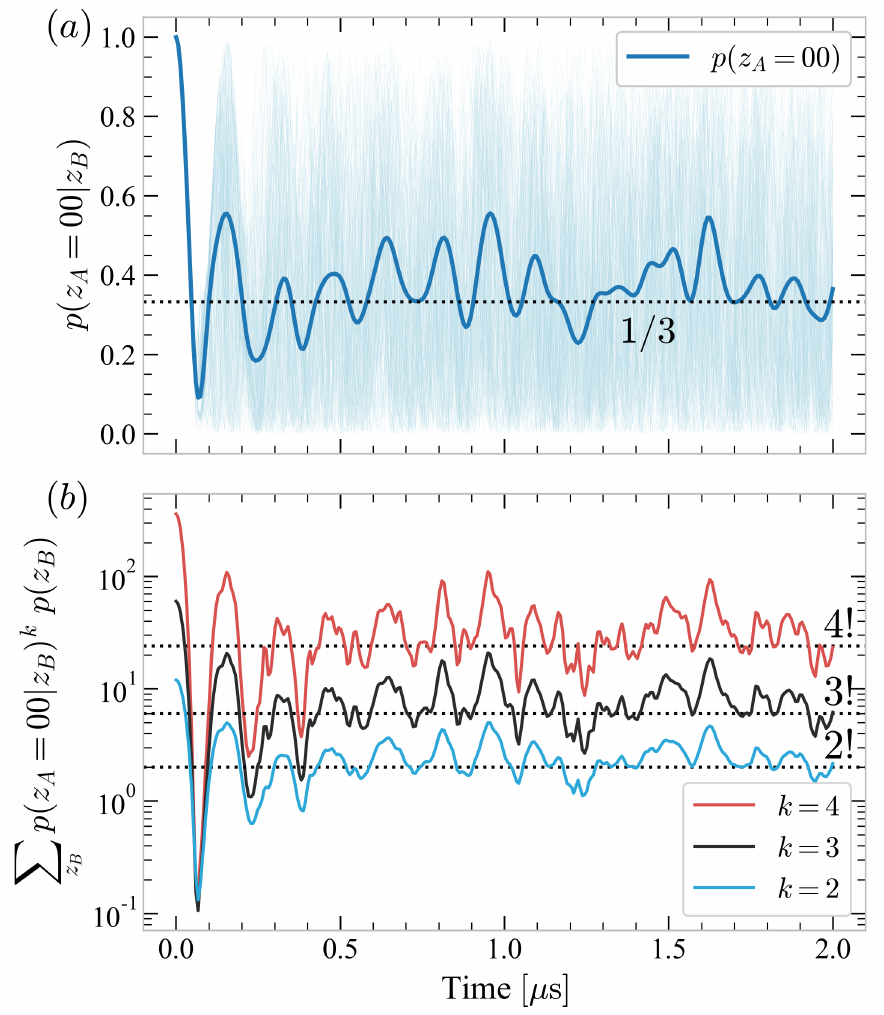}
		\caption{(a) Numerical simulation of the Rydberg dynamics with partition size $N_A=2$ and system size $N=10$: Conditional probabilities $p(z_A=00|z_B)$ are plotted as a function of time for finding A in the ground state given a measurement $z_B$. Lighter lines correspond to the plot of $p(z_A=00|z_B)$ over time for each of the possible $z_B$ outcomes, while the oscillating bold blue line represents $p(z_A=00)$ which reaches a steady state value of  $0.33$. (b) Plot of the convergence of higher moments of the conditional probability distribution $p(z_A=00|z_B)$ for numerically generated measurements with system size $N_A=2$ and subsystem $B$ size $N_B=8$, re-scaled by $D_A \times (D_A + 1) \hdots   (D_A + k-1)$ as indicated in the main text.}
		\label{dynamics}
	\end{figure}

	For the numerical simulation, we consider a system of $N=10$ atoms,  partitioned into two subsystems $A$ and $B$ with $N_A=2$ and $N_B=8$ in Fig.~\ref{fig:system-diagram}.
	The many-body system is initialized  in the ground state $|ggg\hdots\rangle$, which has zero energy expectation value with respect to $\hat H$. The initial state evolves under $\hat H$ to a final state $|\Psi(t)\rangle = e^{-\frac{i}{\hbar} \hat H t}|g\hdots g\rangle$ at time $t$, which is well before any decoherent process, such as spontaneous decay of the Rydberg state  become significant. A natural consequence of the interacting many-body Rydberg system is the transition from the initially ordered state to chaotic  dynamics, characterized by the extensive occupation of a broad manifold of eigenstates across the Hilbert space. This is depicted in the plot of the conditional probabilities shown in Fig. \ref{dynamics}. The evolved state of the many-body system $|\Psi(t)\rangle$ at time $t$ is expressed as $\sum_{z_B} \sqrt{p(z_B)}|\Psi_A(z_B)\rangle \otimes |z_B\rangle$ which provides the data set for the ensemble and is used for characterization in this work. For the numerical implementation of the simulator, we closely follow the experimental setup described in Ref.~\cite{Choi2021}, which serves as our benchmarking platform, using the following parameters: lattice spacing $a=3.3\ \mathrm{\mu m}$ apart, $C_6 = 126\ \mathrm{GHz}\ \mathrm{\mu m^6}$, $\Delta/(2\pi) = 0.9\ \mathrm{MHz}$ and Rabi frequency $\Omega/(2\pi) = 4.7\ \mathrm{MHz}$ \cite{Choi2021}. Practical implementation of measurement in randomly chosen basis with single qubit rotations in Rydberg systems can be achieved in current experiments \cite{Chen_2023, Notarnicola_2023, Guillaume2024}.
	
		\begin{figure*}[t!]
		\centering
		\includegraphics[width=\linewidth]{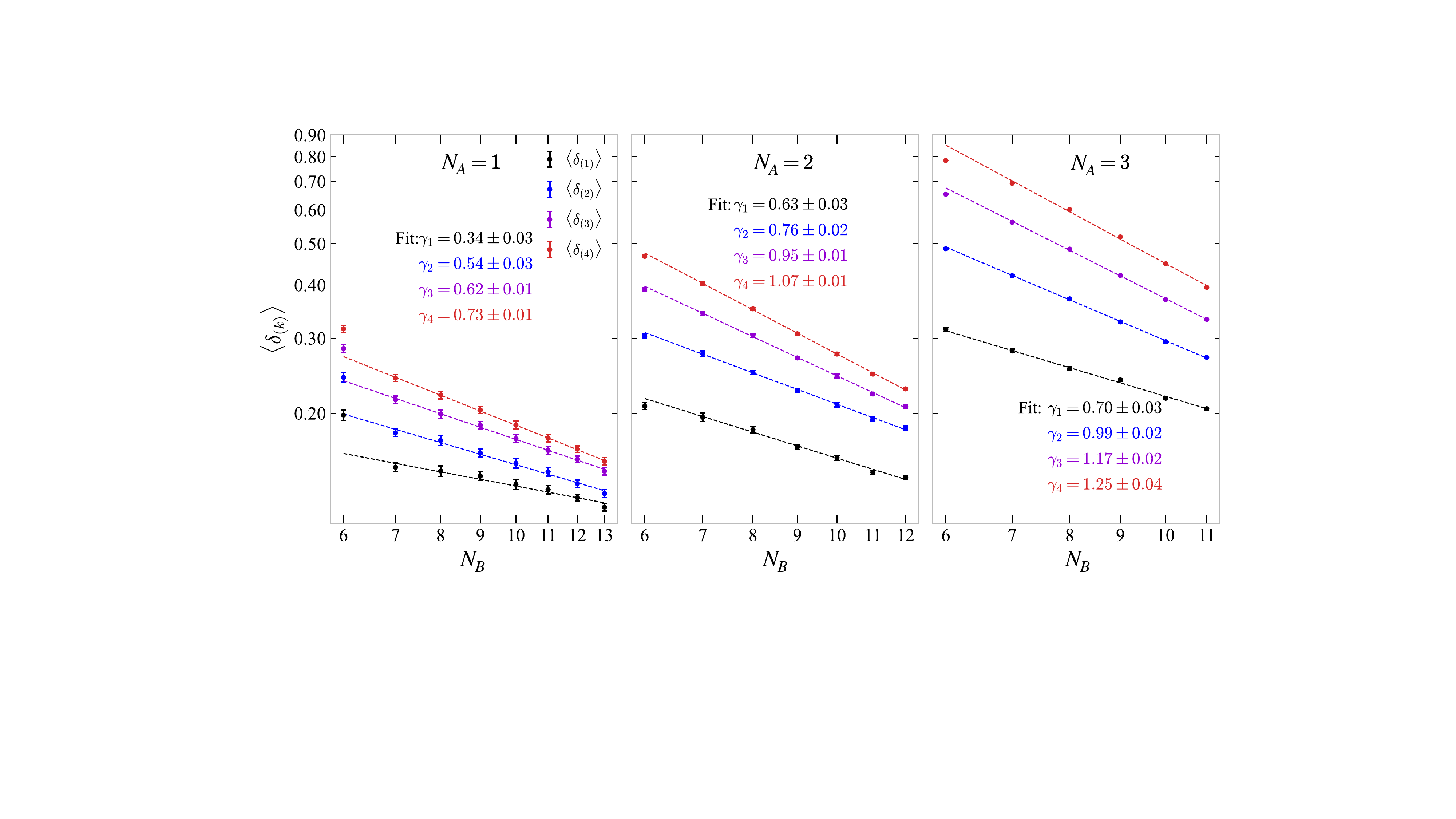}
		\caption{Shows the time average and standard deviation of the value of the trace distance in the steady state ($y$ axis) as a function of different subsystem $B$ sizes ($x$ axis). The system considered has a total size of $N=14$ with varying subsystem and subsystem $B$ sizes. The expected scaling law of the trace distance is established, where the $\gamma_k$ values are shown in the panels corresponding to different quantum state k-designs.}
		\label{scaling}
	\end{figure*}



Now, whether a chaotic final state can provide a projective ensemble of states that statistically represent a quantum state $k$-design can be addressed in two different ways. Firstly, a random ensemble of states is known to possess certain universal statistical properties such as the steady state behavior of the marginal probability distributions $p(r_A=z_A)$ (see Eq.~\ref{marginal} but now in $Z$ basis) at long times \cite{Cotler2021}. Fig.~\ref{dynamics} ascertains the expected convergence of the first moment (mean) to $1/D_A(=1/3)$ and subsequent higher moments of the conditional probability distributions, which are defined as $\sum_{z_B} p(z_B) p(z_A=00|z_B)^{k}$ to values $k!/(D_A \times (D_A + 1) \hdots (D_A + k-1))$ at long times. The curves in (b) are re-scaled by a factor ($D_A \times (D_A + 1) \hdots   (D_A + k-1)$), thus approaches $k!$. $D_A$ is the dimensionality of the Hilbert space for the system in consideration. For a subsystem of two qubits, the blockade interaction prohibits Rydberg excitation next to each other, thereby resulting in $D_A=3$ possible states i.e.~ $z_A\in\{\ket{00},\ket{01},\ket{10}\}$. 
	Although Figs.~\ref{dynamics}(a-b) were done for measurements in $Z$ basis, the saturation of conditional probabilities at long times holds true independent of the basis chosen for projective measurements. The saturation of conditional probabilities with projective measurements in the $Z$ basis has been verified  in the recent experiment \cite{Choi2021} and is the first signature that the ensemble of states is approaching a Haar-random distribution. 
	\begin{figure}[t]
		\centering
		\includegraphics[width=0.99\linewidth]{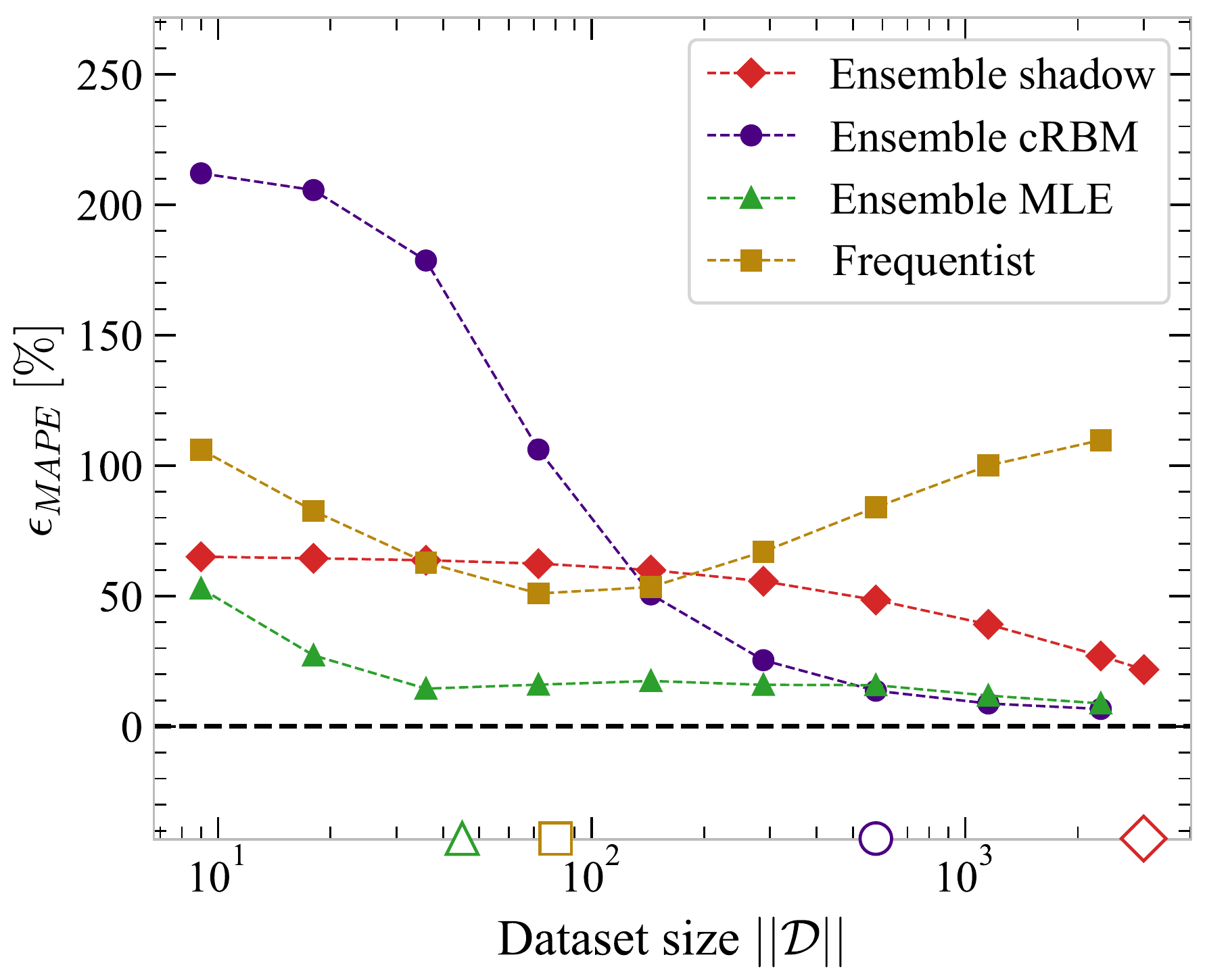}
		\caption{Plot of the mean absolute percentage error as defined in Eq.~\ref{mre} for $k=2$ with each of the four methods. For statistical robustness, the mean percentage error  is evaluated by averaging across 21 equally spaced time steps in the thermalized regime from $0.4\ \mathrm{\mu s}$ and $1.4\ \mathrm{\mu s}$ and across 10 repetitions of the numerical experiment for every time step. The sample size on the $x$ axis corresponds to the number of measurements made in each repetition, and the empty markers correspond to the =optimal dataset sizes (Shadow: 3000, cRBM: 576,  MLE: 45, Frequentist: 80) in the range explored.}.
		\label{datasetsize}
	\end{figure}

	\begin{figure}[t!]
		\centering
		\includegraphics[width=0.9\linewidth]{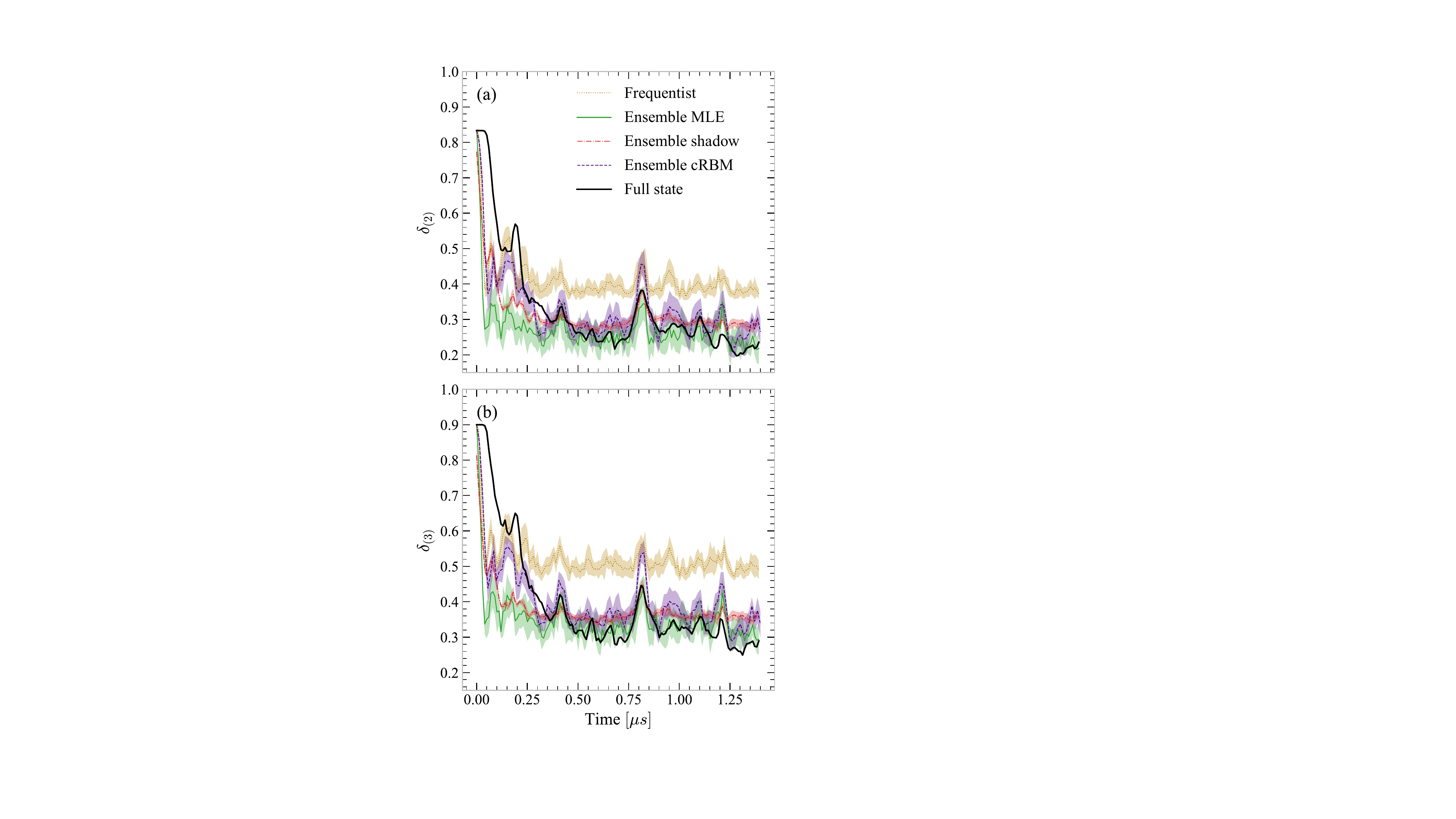}
		\caption{The figure shows trace distances estimated for 2-design ($\delta_{(2)}$) and 3-design ($\delta_{(3)}$) using different algorithms as a function of time. Each algorithm was performed on a dataset of size marked on the $x$-axis of Fig.~\ref{datasetsize}. For statistical robustness, at each time the simulated experiment was repeated 10 times. The mean reconstructed trace distance is shown with a line and the standard deviation is shown as a shaded region. The black line shows the target values, calculated using the simulated time-dependent wave function without approximation.}
		\label{fig:trdist23} 
	\end{figure}
	However, a more direct approach for verifying that a certain ensemble state  forms an approximate quantum $k$-design is to demonstrate that the trace distance $\delta_{(k)}$, which is defined in Eq.~\ref{eq:tr-distance}, approaches zero. To evaluate the trace distance (Eq.~\ref{eq:tr-distance}), we first construct the ensemble in Eq.~\ref{eos} and its $k$-th moment in Eq.~\ref{eq:moment-ens} using the state vector $|\Psi_A(z_B)\rangle$ and probabilities $p(z_B)$, which are derived from the time-evolution of Eq.~\ref{eq:hamiltonian} followed by the simulated projective measurements on subsystem $B$. The steady-state value of the trace distance, $\langle \delta_{(k)} \rangle$, is then obtained by averaging $\delta_{(k)}(t)$ over 130 uniformly spaced time points within the interval $(1.5, 5.0) \mu\mathrm{s}$. Figure \ref{scaling} shows the numerical results for the emergence of approximate $k$-designs in subsystem sizes ($N_A=$ 1, 2 and 3) as a function of the subsystem $B$ size $N_B$. In this analysis, we consider various possible ways of bi-partitioning the system, demonstrating that the characterization of a $k$-design is independent of the specific choice of subsystem $A$. Furthermore, the dependence of $\langle \delta_{(k)} \rangle$ on $N_B$ follows a simple scaling law, $\langle \delta_{(k)} \rangle  \propto \frac{1}{(N_B)^{\gamma_k}}$, consistent with the recent numerical observations of Ref.~\cite{Cotler2021}
	The exponents $\gamma_k$  are obtained using linear fitting the data on a log-log scale. Moreover, a comparison across the three subplots reveals that $\gamma_k$ increases with $N_A$ for a given $k$, indicating that, when subsystem $B$ is sufficiently large, larger subsystem sizes $A$ yield more uniform approximate designs. However, evaluating the trace distance generally requires full knowledge of the quantum state, which is experimentally exhaustive. To overcome this limitation, we demonstrate in the next section how statistical learning models can be employed to estimate $\Psi_A (z_B)$, and thereby the trace distance, from a limited number of measurements.
	
	\section{Results}
	The need for accurate quantum state tomography for ensemble states is expensive in measurement cycles. It has been proven that for an $N$-body quantum system, the classical data obtained from randomized measurements can approximate all reduced density matrices with fixed number of spins up to $\epsilon$ accuracy with $log(N)/(\epsilon^2)$ measurements \cite{Vermersch,Elben}. Following this up with efficient post-processing of the classical data with methods such as maximum likelihood optimization or RBM can potentially allow future experiments to estimate trace distances between the fitted many-body wave function and Haar-random distribution. The results of our approach are compared against the predictions of shadow tomography, where the density matrix is reconstructed using a computationally more efficient $U_1$ statistics with negligible bias for the sample size considered here, and is discussed in more detail in Appendix \ref{shadowtom}.

	Figure \ref{datasetsize} compares the performance of  different approaches in terms of the trace distance with respect to the size of the measurement dataset. Estimates calculated from each of the different methods are compared to the exact trace distance $\delta^e_{(k)}$ calculated numerically using the true state vector $|\Psi(t)\rangle$ obtained by numerically solving the dynamics of the Rydberg system. The value of $\delta^e_{(k)}$ will in general be non-zero, since the system we are studying produces only \textit{approximate} state designs. The comparison between $\delta^{(e)}_{(k)}$ and $\delta^{(m)}_{(k)}$ for each method $m$ is performed through the mean absolute error, defined as 
	\begin{equation}\label{mre}
		\epsilon_{MAPE} = \frac{1}{(M\times T)}\sum_{j=1}^{M} \sum^{T}_{i=1} \left |\frac{{\delta^{m,j}_{(k)}(t_i)} - \delta^e_{(k)}(t_i)}{\delta^e_{(k)}(t_i)}\right | ,
	\end{equation}
where $\delta^{m, j}_{(k)}(t_i)$ is the value of the trace distance obtained using Eq.~\ref{eq:tr-distance} for dataset-based method $m$ evaluated at time step $t_i$. The label $j$ indexes the repetition of the numerical experiment. Thus, Fig.~\ref{datasetsize} shows the efficacy of the chosen method $m$ in characterizing the ensemble of states as a quantum $k$-design with respect to the number of measurements needed for $k=2$. For statistical robustness,  the mean of the relative error is calculated over $T=21$ time steps and $M=10$ repetitions of the sampling process. In Fig.~\ref{datasetsize},  we find that the best method for characterizing the ensemble of states is the  MLE method, showing a relatively low value of the error $\epsilon_{MAPE}$ across the range of dataset sizes studied in this work, including in the low sample regime. For $||\mathcal{D}||>1000$ simulated measurements, $\epsilon_{MAPE}$ converges to zero. The cRBM approach shows poor results in the low sample regime but improves rapidly as more measurements become available ($||\mathcal{D}||>1000$). cRBM performs much worse than  MLE in the low sampling regime because the actual number of samples that can be used for training the cRBM is much lower than the size of the available dataset $\mathcal{D}(z_B)$. This is related to a technical constraint where the algorithm requires at least one measurement to be in the $Z$ basis (which is hardwired in the software implementation \cite{Beach2019}) for subsystem $A$, which is sometimes missed in the random measurement scheme. Removing this constraint in future implementations of the software will potentially help in making better predictions. Shadow tomography performs better than cRBM in the low sampling regime but its error $\epsilon_{MAPE}$ converges to zero  slower than cRBM and  MLE. It appears that reconstructing the density matrix directly from the measurements in randomly chosen bases, as performed in shadow tomography (refer to \ref{shadowtom}) requires more measurements to achieve the same level of accuracy in characterizing quantum $k$-designs compared to maximum-likelihood method. Lastly, the standard frequentist approach exhibits a consistent overestimation of the trace distance throughout the investigated dataset sizes, with no significant improvement with increasing sample size. This is because in the frequentist approach we have only $Z$ basis measurements for subsystem $A$, as is typical in conventional Rydberg experiments, which implies that it does a poor job of capturing the full information of the state function. Although all approaches perform better in the high measurement regime, the  MLE method has better accuracy and it is a few orders of magnitudes faster to train. The stars on the $x$-axis indicate the minimum number of simulated measurements needed at each time step in order to get the best possible result for each of the algorithms within the range of dataset sizes analyzed. These are also summarized in Table \ref{table_1} and are used to compare each of the methods in Fig. \ref{fig:trdist23}.

		\begin{table*}[htbp]
	\centering
	
	\begin{tabular}{|l|c|c|c|c|}
		\hline
		\textbf{Feature} & \textbf{Frequentist} & \textbf{Max. Likelihood} & \textbf{cRBM} & \textbf{Shadow Tomography} \\
		\hline
		Measurement basis ($A$) & Z-basis & Random basis & Random basis & Random basis \\
		\hline
		State representation  & $\sum p_m |m\rangle$ & $ \sum_{z'_A} \sqrt{\frac{p(z'_A,z_B)}{p(z_B)}}|z'_A\rangle$  & $\sqrt{p_{\theta}(z)}\ e^{i\phi_{\mu}(z)}$ & $ \hat{\rho}_A^{(n)}= \bigotimes_i \left( 3 |a_i^{(n)}\rangle\langle a_i^{(n)}| - I\right)$ \\
		$\ket{\Psi_A(z_B)}$& $p_m \rightarrow$ prob.~of $\ket{m}$  & $p(z'_A,z_B)\rightarrow$ Joint prob. & $(p_{\theta},\phi_{\mu})\rightarrow$ (Amp., Phase) & $\ket{a_i^{(n)}}\rightarrow$ Measurement$_{i-qubit}^{n-shot}$ \\
		\hline
		Estimation error & --- & $1/||\mathcal{D}||$ \cite{gill2000state} & $1/\sqrt{||\mathcal{D}||}$ \cite{Torlai2019paper} & $1/\sqrt{||\mathcal{D}||}$ \cite{Huang} \\
		\hline
		Optimal dataset size& --- & $45$ & $576$ &  $3000$\\
		\hline
	\end{tabular}
	\caption{Summary of different approaches for characterizing quantum state ensembles. The key methodological features of the frequentist, maximum-likelihood (MLE), complex Restricted Boltzmann Machine (cRBM), and shadow-tomography approaches are shown in top three rows. Each method differs in the measurement basis, statistical representation of the state, and the scaling of the sampling requirements. See text and \Cref{estimators} for more details. The optimal dataset size for characterizing quantum state ensembles with each approach is obtained from Fig.~\ref{datasetsize}.}
	\label{table_1}
\end{table*}
	
Figure \ref{fig:trdist23} plots the trace distances $\delta_{(2)}$ and $\delta_{(3)}$ as a function of time in order to verify the quantum state 2- and 3-designs respectively. The black line is the trace distance calculated using $|\Psi_A(t)\rangle$ which corresponds to the case where there is complete knowledge of the state. The other lines show the estimations obtained with each of the three algorithms using simulated measurements. The algorithms were applied to $M=10$ independent simulated datasets at each time point, and the plot shows the mean (solid line) and standard deviation (shaded area) across the repetitions. In the initial times, all four algorithms display severe underestimation of the $\delta_{(2,3)}$ value. However, in the steady-state regime (after about $0.25\ \mathrm{\mu s}$), the estimates of the trace distance from the MLE, cRBM and shadow methods approach the true value (solid black line), clearly outperforming the frequentist approach. Apart from predicting the average value for $\delta_{(k)}(t)$, it is interesting to find that these  four methods also capture some of the fluctuations in the trace distance that is seen using the exact method. The shadow tomography is not performing as well as MLE, despite the fact that the classical shadow method has a two orders of magnitude larger number of measurements for the trace distance estimation compared to  MLE (also see Fig.~\ref{datasetsize}). We suggest an explanation for this which is that cRBM and  MLE attempt the reconstruction of a state vector, whereas shadow tomography reconstructs a density matrix. The increased number of parameters which need to be estimated may be responsible for the higher number of samples required for the characterization of the system using the classical shadow method.

\section{Discussion}

The complexity of large quantum systems makes them difficult to characterize accurately and full state tomography is inefficient as it requires an exponentially large number of experimental runs which also amounts to exponentially large amount of data to process. Improving the efficiency of quantum state tomography, in particular for large systems is a constant challenge and an active area of research \cite{Cramer,LanyonTomo}. Characterizing the randomness of an ensemble of states is no exception to this plight. Full characterization of a quantum state density matrix in a Hilbert space of dimensionality $D$ necessitates at least the preparation of $O(D^2)$ independent copies of the system~\cite{o2016efficient, haah2016sample}. However, the Hilbert space of $N$ interacting spins such as that of the Rydberg setup has dimensionality $D=2^N$, in which case the full characterization of a quantum state density matrix $\rho$ will require  $O(D^2) = O(2^{2N})$ independent copies of the system, which is still an exhaustively large number of copies. 
	
Often a far less complete description of the state is adequate in certain cases such that the number of experimental runs and the amount of data needed is drastically reduced. The key idea is to translate the information about the many-body state of a system into classical data by performing measurements in the random bases. We sacrifice some information to make characterization feasible, aiming to strike a balance between data requirements and accuracy.  In this regard, classical shadow tomography has a relatively reduced scaling to $O(\log^4(\mathcal{M})\log(D))$~\cite{Aaronson2018} where $M$ are the two-outcome measurements \cite{Aaronson2018}.  Measurements in randomized basis can reduce this scaling further to  $O(\log(\mathcal{M}))$ independent copies \cite{huang2020predicting, elben2023randomized} and is at the heart of the modern classical shadow methods~\cite{huang2020predicting}. However, depending on the many-body quantum state that is being characterized, relevant information can be lost leading to an inaccurate description of the quantum state. In such cases, statistical learning methods can be used to make accurate predictions with low estimation error as outlined in Table \ref{table_1} and, in certain cases, more efficient  predictions of the quantum state, as demonstrated in this work on characterization of emergent random ensembles of pure states.
	
	This work uniquely accomplishes the efficient characterization of ensembles of random states by combining the benefits of random basis state measurements and statistical learning methods. The ability to identify information in reduced data sets that would otherwise be missed by naive observations is one of the appealing features of this kind of learning \cite{Masahiro}. In contrast to the shadow tomography, which requires more than $10^3$ measurements to converge even with efficient $U_1$ statistics, MLE ($\sim50$) and cRBM ($\sim500$) requires a significantly fewer measurements. While both methods,  MLE and cRBM yield  more accurate estimations, the current implementations of cRBM are computationally more time-consuming by orders of magnitude when compared to the MLE approach. This may change in the future with better and more flexible versions of cRBM routines, however, currently, this work indicates that MLE offers the most practical balance between accuracy and efficiency, and it is applicable to a broad class of quantum simulators that realize spin models.
	
	Future works can involve noise modeling that takes into account imperfections in measurements, perhaps using Bayesian methods \cite{mukherjee2020preparation,Sauvage,Mukherjee} and could make comparisons between the approach used in this work with other similar methods such as shadow estimation \cite{Huang}, as well as the case of mixed-states occurring due to non-unitary processes in the dynamics.  Moreover, our findings resonate with complexity-theoretic arguments such as anti-concentration theorems, which establish the necessity of approximate designs for demonstrating quantum speedups and validating randomness in near-term devices~\cite{hangleiter2018anticoncentration}.
	
	\acknowledgments
	
	RM acknowledges support from the U.S. National Institute of Standards and Technology (NIST) through the CIPP program under Award No. 60NANB24D218.

	\appendix	
	\section{Expression for $k^\text{th}$ moment for random Haar ensemble}\label{AnaHaar}
	Consider a Hilbert space with dimensionality $d$, then the Haar ensemble is defined as the continuous random probability distribution over pure states defined over that space. Using the Schur-Weyl duality, an analytical expression to calculate the  $k^{th}$ moment of the Haar ensemble \cite{Harrow2018} is given as 
	\begin{equation}\label{kHaar}
		\hat\rho_{Haar}^{(k)} = \frac{\sum_{\pi \in S_k} \hat P_d(\pi)}{d (d+1) \dots (d + k -1)}.
	\end{equation}
	The sum in the expression above iterates through the elements $\pi$ of the permutation group $S_k$. These represent all the ways of rearranging an array of $k$ distinct objects. For example, we explicitly report the elements of the group $S_3$ below, which are needed to construct the third moment of the ensemble $\rho_{Haar}^{(3)}$.
	\begin{align}
		&\pi_1: \left\{\begin{matrix}
			\pi_1(1) = 1\\
			\pi_1(2) = 2\\
			\pi_1(3) = 3
		\end{matrix}\right.
		& \pi_2: \left\{\begin{matrix}
			\pi_2(1) = 2\\
			\pi_2(2) = 1\\
			\pi_2(3) = 3
		\end{matrix}\right.\\[2mm]
		& \pi_3: \left\{\begin{matrix}
			\pi_3(1) = 3\\
			\pi_3(2) = 2\\
			\pi_3(3) = 1
		\end{matrix}\right.
		&\pi_4: \left\{\begin{matrix}
			\pi_4(1) = 1\\
			\pi_4(2) = 3\\
			\pi_4(3) = 2
		\end{matrix}\right.\\[2mm]
		& \pi_5: \left\{\begin{matrix}
			\pi_5(1) = 3\\
			\pi_5(2) = 1\\
			\pi_5(3) = 2
		\end{matrix}\right.
		& \pi_6: \left\{\begin{matrix}
			\pi_6(1) = 2\\
			\pi_6(2) = 3\\
			\pi_6(3) = 1
		\end{matrix}\right.
	\end{align}
	The operator $\hat P_d (\pi)$ is like a SWAP operator that acts on $k$ copies of $d$ dimensional qudits and is defined as follows \cite{Harrow2018},
	\begin{equation}
		\hat P_d(\pi) = \sum_{i_1,\dots, i_k \in [1,2,\dots,d]} |i_{\pi(1)},  \dots, i_{\pi(k)}\rangle \langle i_1, \dots, i_k| .
	\end{equation}

	\section{Simulating measurement data}\label{measuredata}
\begin{figure*}[t!]
		\centering
		\includegraphics[width=0.99\linewidth]{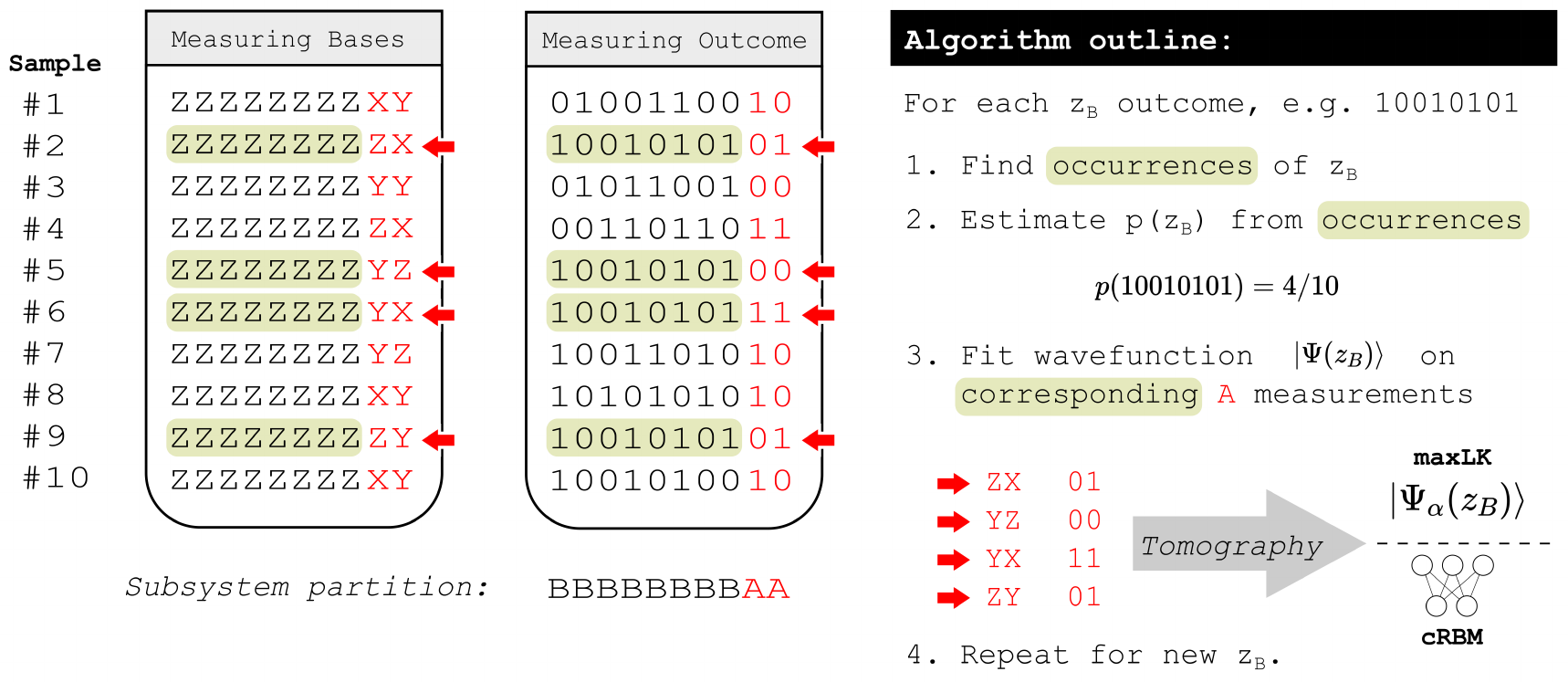}
		\caption{The figure illustrates the scheme for generating the simulated measurement datasets $\mathcal D (z_B)$. The numerically generated bitstrings are collected and sorted in sub-sets containing the common $z_B$. A probability is assigned to each unique $z_B$ string by counting the number of occurrences of that bitstring in the dataset. The corresponding measurements of $A$ associated to a particular $z_B$ determine the state $|\Psi_A(z_B)\rangle$. The tuple $(p(z_B), |\Psi(z_B)\rangle)$ accordingly defines the ensemble state $\mathcal{E}$.}
		\label{enhanced-measurement}
	\end{figure*}

	Repeated measurements are taken to obtain statistics by randomly sampling the measurement directions between the three mutually orthogonal directions. For spin systems, and in particular for Rydberg systems, this can be experimentally implemented by randomly rotating each spin using a unitary operator followed by a state measurement in the z basis \cite{brydges2019probing}. The data set $\mathcal{D}$ of size $||\mathcal{D}||$ is constructed by measuring the subsystem $B$ in the $z$ basis and the system $A$ in all bases. It is subsequently split into subsets that share the same outcome $z_B$ of the subsystem $B$ measurement, which we label as $\mathcal{D}(z_B)$. Each $\mathcal D (z_B)$ contains multiple simulated measurements of subsystem $A$ for a particular $z_B$ as shown in Fig.~\ref{enhanced-measurement}. The $i$th measurement in $A$ is denoted by $|r_A^{(i)}(z_B)\rangle_\mathbf{b} = |b_{A,1}^{(i)}\rangle \otimes \hdots \otimes |b_{A,N_A}^{(i)}\rangle$ where $|b_{A,j}^{(i)}\rangle  \in |\{0,1\}\rangle_{\{X, Y, Z\}}$ are the eigenstates of the $\sigma_{b_j}$ Pauli operator with eigenvalue $\pm 1$. It should be noted that the amount of data available to train the state $|\Psi(z_B)\rangle$ for a particular $z_B$ fluctuates for each run of numerically simulated measurements. In general, $z_B$ bitstrings with a higher probability $p(z_b)$ are more likely to be sampled and the size of the corresponding dataset $\mathcal{D}(z_B)$ will be on average larger. This also implies that subsets $\mathcal{D}(z_B)$ with low $p(z_b)$ may have measurements for $A$ that are missing in certain directions. For example, in the illustrative set of measurements provided in Fig.~\ref{enhanced-measurement}, the dataset $\mathcal{D}(10010101)$ contains measurements of subsystem $A$ in the basis $\{ZX, YZ, YX, ZY\}$ for a given outcome $z_B$. Here the first spin for $A$ was never measured in the $X$ basis which may then affect the quality of the training. However, the states with particularly low $p(z_B)$ do not contribute significantly to the $\rho_\mathcal{E}^{(k)}$ thereby reducing the impact of inaccurate state characterization due to the limited data.  
	
	\section{Estimators for random ensemble state}\label{estimators}
	In this section, we outline the details of the four statistical learning methods employed in this work, namely (i) complex restricted Boltzmann machines (cRBM), (ii) Maximum likelihood estimation, (iii) frequentist approach and (iv) Shadow tomography. 
	\begin{figure}[t]
		\centering
		\includegraphics[width=0.99\linewidth]{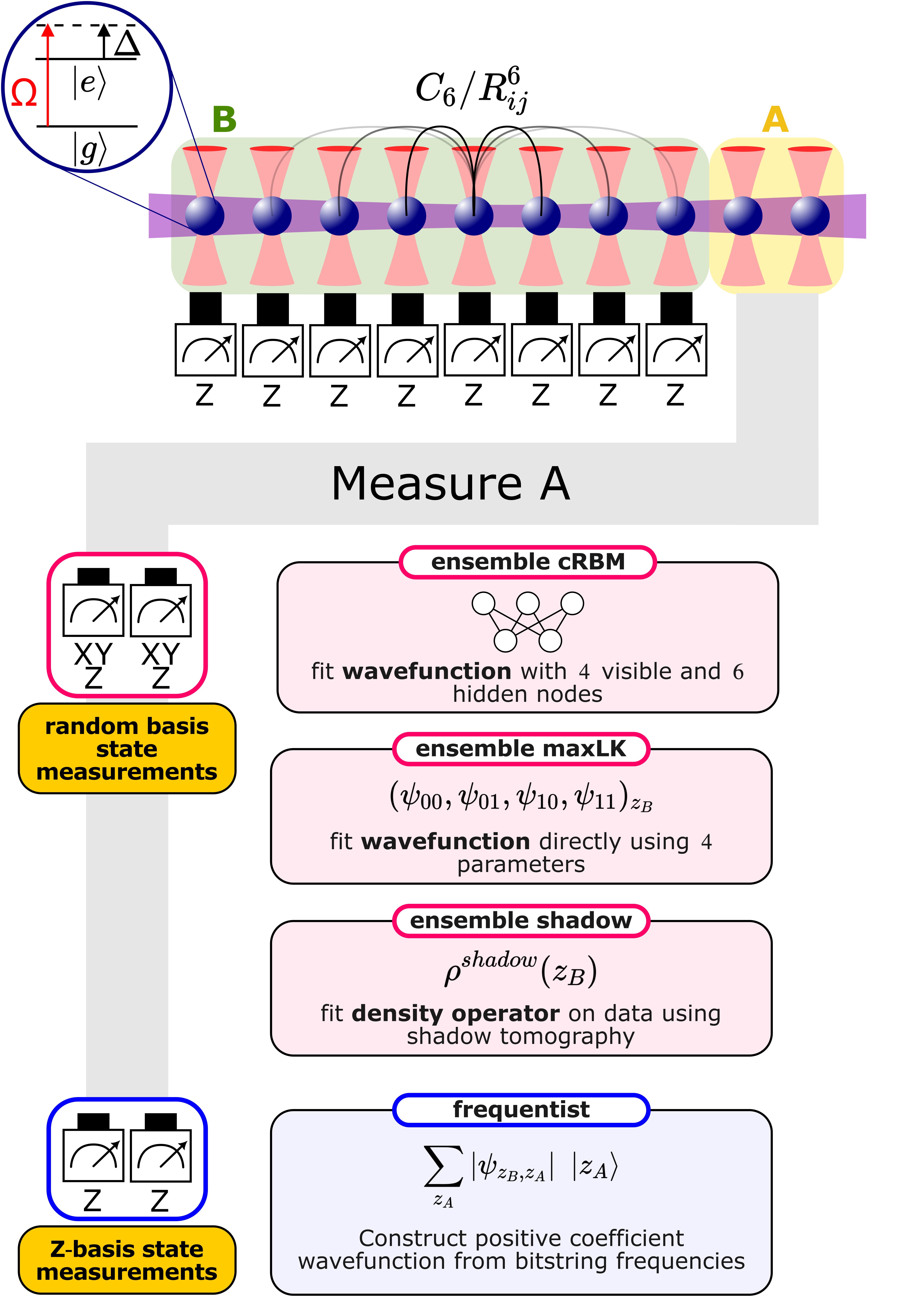}
		\caption{This figure schematically summarizes the 4 approaches used to verify approximate quantum k-design. Numerical simulations provide measurement data for subsystem $B$ in $Z$ basis with probabilities $p(z_B)$ for all 4 methods. In the case of ensemble maxLK, cRBM and shadow ensemble approaches, the measurement basis for subsystem $A$ is chosen randomly in one of the three directions. Applying maxLK when system $A$ is measured only in the $Z$ basis becomes equivalent to the frequentist method.}
		\label{fig:algorithms}
	\end{figure} 
	
	\subsection{Complex restricted Boltzmann machines}
\begin{figure}[t!]
		\centering
		\includegraphics[width=0.95\linewidth]{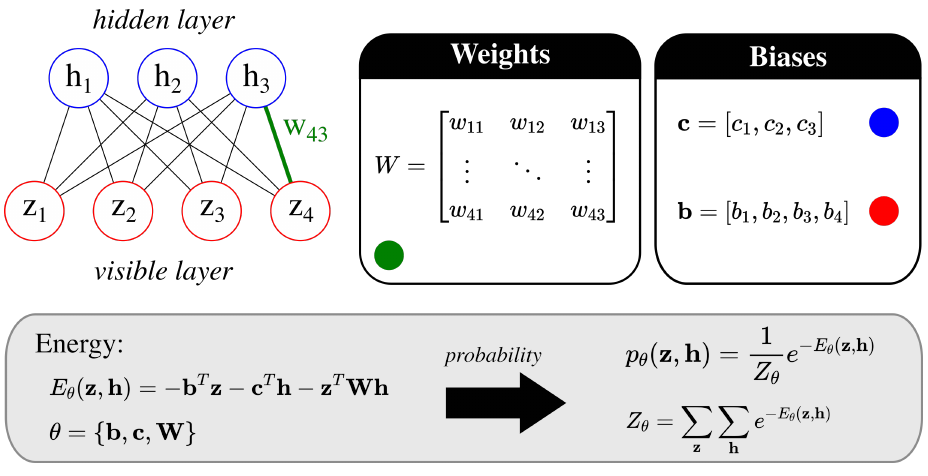}
		\caption{Network diagram of RBM. The network is divided into a hidden and visible layer, each one consisting of nodes $h$ and $z$ that take binary values 0 and 1. RBM is parameterized by $\theta$ which groups the weights associated to the interlayer connections and biases associated to the hidden and visible nodes. An energy $E_\theta(\bf \, \bf h)$ can be calculated for each network configuration. The energy allows defining a probability $p_\theta(\bf \, \bf h)$ that is inspired by the Boltzmann distribution. $Z_\theta$ is the normalizing partition function.}
		\label{fig:rbm}
	\end{figure}

	A restricted Boltzmann machine (RBM) is a type of bipartite neural network with two layers which are referred to as visible and hidden layers, as shown in Fig. \ref{fig:rbm}. It is used to learn the probability distribution over binary data in an unsupervised approach. The network shows no intra-layer connections but just interlayer ones, thus motivating the name \textit{restricted}. The nodes in the visible and hidden layers have bias vectors $\mathbf b$ and $\mathbf c$ associated to them, whereas a matrix $\mathbf W$  provides weights for the connections.  The nodes $z_i$ and $h_j$ can only take binary values 0 or 1 and are thus used to represent a probability distribution $p(\mathbf z)$, where $\mathbf z$ is a vector of $N$ binary entries $\{0,1\}$ i.e. $\mathbf z \in \{0,1\}^N$. The state of the RBM is completely defined using the two vectors:
	\begin{subequations}
		\begin{align}
			\mathbf z &= (z_1,z_2,...,z_V)\qquad &z_i \in \{0,1\}\\
			\mathbf h &= (h_1, h_2, ..., h_H)\qquad &h_j \in \{0,1\}
		\end{align}
	\end{subequations}
	where $V$ and $H$ are the total numbers of nodes in the visible and hidden layers respectively. Each state of the RBM has an associated energy given by
	\begin{equation}
		E_{\theta}(\bf{z}, \bf{h})=-\bf{b}^{T} \bf{z}-\bf{c}^{T} \bf{h}-\bf{z}^{T} {W \bf h},
	\end{equation}
	which is parametrized by the weights and biases of the model $\theta = \{\bf b, \bf c, W\}$. The probability distribution $p_\theta(\mathbf{z}, \mathbf{h})$ (chosen to be a Boltzmann distribution)  and the partition function $Z_\theta(\bf z, \bf h)$ associated to the state energy are given as
	\begin{align}
		&p_\theta(\mathbf{z}, \mathbf{h}) = \frac{1}{Z_\theta}e^{-E_\theta(\mathbf{z}, \mathbf{h})} , \\
		&Z_\theta = \sum_{\mathbf z}\sum_{\bf h}e^{-E_\theta(\bf z, \bf h)} .
		\label{eq:partition-function}
	\end{align}
	In our work, we are interested in the probability $p_\theta(\bf z)$, which is the marginal of $p_\theta(\bf z, \bf h)$ and has the following analytical expression,
	\begin{align}\label{eq:effective-energy}
		p_\theta(\bf{z}) &= \sum_{\bf h'} p_\theta(\bf z, \bf h') = \frac{1}{Z_\theta} e^{-\mathcal{E}_\theta(\bf z)}, \\
		\mathcal{E}_{\theta}(\bf{z})&=-\bf{b}^{T} \bf{z}-\sum_{j=1}^{H} \ln \left[1+\exp \left(c_{j}+\sum_{i=1}^{V} W_{i j} z_{i}\right)\right] .
	\end{align}
	Using this formalism, a particular state $\bm{z} = {010...1}$ is associated with a probability $p_\theta(z)$ that is parametrized by $\theta = \{\bf b, \bf c, W\}$. The goal is to find the optimal parameters $\theta$ that give the approximation $p_\theta(\bf z) \approx q(\bf z)$ given some data
	\begin{equation}
		\mathcal D = [z^{(1)},z^{(2)},\dots, z^{(D)}], \qquad \mathrm{where}\ z^{(i)}\in \{0,1\}^{V} \label{eq:cl-rbm-data}
	\end{equation}
	sampled according to the target distribution $q(\bf z)$ which the data are sampled from. This is referred to as training of the RBM with respect to the data.\\
	
	\textbf{Training the RBM with data:} The commonly used cost function that needs to be minimized while training the RBM is the Kullback-Leibler divergence \cite{Torlai2019paper,Melko2019} which is defined as
	\begin{equation}
		C_\theta = \sum_z q(z) \log{\frac{q(z)}{p_\theta(z)}}  = \sum_z q(z) \log{q(z)} - q(z) \log{p_\theta}(z), 
		\label{eq:cost}
	\end{equation}
	where the term $\sum_z q(z)\log{p_\theta(z)} = \langle \log{p_\theta} \rangle_q$ is the averaged value of $\log{p_\theta(z)}$ with respect to $q(z)$. Given the function $C_\theta$, the parameters are updated to minimize this using the following rule:
	\begin{equation}
		\theta \leftarrow \theta - \eta \nabla_\theta C_\theta
	\end{equation}
	Here $\eta$ is the \textit{learning rate}, and it affects the step size of the parameter update. The dependence of the cost function on the parameters is captured by the following expression \cite{Torlai2019paper,Melko2019},
	\begin{equation}
		\nabla_\theta C_\theta \approx \langle \nabla_\theta \mathcal E_\theta \rangle_{\mathcal D} - \langle \nabla_\theta \mathcal E_\theta \rangle_{p_\theta} .
	\end{equation}
	The first term is often called the \textit{positive phase}, and it represents the average of the gradient of $\mathcal{E}_\theta$ (cfr. \ref{eq:effective-energy}) evaluated at the values of $z$ in the dataset $\mathcal D$. The second term is the \textit{negative phase}, and it represents the average of the gradient of $\mathcal{E}_\theta$ taken over the learned probability distribution $p_\theta(z)$. This last term can be computationally expensive to evaluate during training. Indeed, the expression for $p_\theta(z)$ involves the calculation of the partition function $Z_\theta$ in \ref{eq:partition-function} and is a task that scales like $\mathcal O(2^H \times 2^V)$, which becomes intractable for large systems. The standard approach to calculate the value $\langle \nabla_\theta \mathcal E_\theta \rangle_{p_\theta}$ is using an algorithm called \textit{contrastive divergence}. This algorithm allows to sample $z$ bitstrings according to $p_\theta$ without calculating the partition function of the system explicitly. A detailed explanation of the optimization algorithm is out of the scope of this work, but it can be found in  \cite{Torlai2019paper,Melko2019}. For a system size of $2$ qubits, the wave function is described as
	
	\begin{equation}\label{ansatz}
		|\Psi\rangle = \psi_{00}|00\rangle + \psi_{01}|01\rangle + \psi_{10}|10\rangle + \psi_{11}|11\rangle
	\end{equation}
	
	and following Born's rule, the probability distribution is given by $p(z)=|\psi_z|^2$. This provides the data that will be used to train the RBM, which in turn implies optimizing the parameters $\theta$ such that the probability $p_\theta(z) \approx p(z)$. Wave functions can be reconstructed using positive real-valued coefficients  $\psi_{z,\theta} := \sqrt{p_\theta(z)}$, but this approach has the disadvantage that phase information is discarded.  A more general approach suggested by Torlai and Melko \cite{Torlai2018Thesis} is to consider two RBMs, one that describes the amplitude of the coefficients $\psi_z$ and the other that describes the phase $\phi(z)$ of the coefficients $\psi_z$. The two RBMs are parameterized with their own weights and biases separately, which are labeled as $\theta$ and $\mu$. They both have associated effective energies $\mathcal{E}_\theta(z)$ and $\mathcal{E}_\mu(z)$ described by Eq.~(\ref{eq:effective-energy}). The coefficients of the wave function are now parameterized as:
	\begin{align}
		\psi_{z,\theta\mu} &= \sqrt{p_{\theta}(z)}\ e^{i\phi_{\mu}(z)} = Z_\theta^{-1/2}\ e^{-\mathcal{E}_\theta(z)/2}\ e^{-i\mathcal{E}_\mu(z)/2} \nonumber \\
		& = Z_\theta^{-1/2}\ e^{-(\mathcal{E}_\theta(z)+i\mathcal{E}_\mu(z))/2} .
	\end{align}
	One can reconstruct the phase $\phi_z$ of the quantum state by measuring along different directions. In this manner, a complex quantum wavefunction can be represented by an RBM and its parameters  $(\theta, \mu)$  can be trained to learn the coefficients $\psi_{z,\theta\mu} \approx \psi_z$.
	

	\subsection{Maximum likelihood method}
	Another way to efficiently estimate $|\Psi_A(z_B)\rangle$ from the data set is to apply a maximum likelihood fitting (MLE). Similar to the previous approach, the coefficients in Eq.~\ref{ansatz} are fitted in a manner that maximizes the \textit{likelihood} of the dataset $\mathcal{D}(z_B)$. This is achieved by minimizing the following cost function \cite{Torlai2018Thesis},
	\begin{equation}\label{eq:cost-fn}
		\mathcal{C}_{\boldsymbol \psi}^{\mathcal{D}(z_B)}=-\frac{1}{\|\mathcal{D}(z_B)\|} \sum_{i=1}^{||\mathcal{D}(z_B)||}  \log \left|\langle \Psi_A(z_B; \bm\psi)    | r^{(i)}_{A} (z_B)\rangle\right|^{2}.
	\end{equation}
	over all the possible vectors of complex coefficients $\bm \psi$ in Eq. \eqref{ansatz}. The optimization is numerically implemented using the Python library NumPy \cite{Myung2003, Harris2020}. Interestingly, if the measurements performed on $A$ are restricted to the $Z$ basis, then minimizing the above cost function is equivalent to implementing the frequentist method which uses the following ansatz:
	\begin{equation}
		|\Psi_A(z_B)\rangle = \sum_{z'_A} \sqrt{p(z'_A|z_B)}|z'_A\rangle =  \sum_{z'_A} \sqrt{\frac{p(z'_A,z_B)}{p(z_B)}}|z'_A\rangle .
	\end{equation}
	The joint probability $p(z_A,z_B)$ is estimated from the outcome frequencies in the data sets. The frequentist analysis is the approach used in the Rydberg experiment \cite{Choi2021}, and served as a benchmark for the two methods proposed in this work.

	\subsection{Equivalence between Maximum-likelihood and Frequentist approaches} \label{MaxFreq}
	In this subsection, we mathematically show that the frequentist approach is just a particular case of the max-likelihood when considering a positive coefficient wave function. There are $M=4$ possible outcomes for a two-qubit wave function $|\Psi\rangle$  that is measured only in the $Z$ basis which are $(00, 01, 10, 11)$. The probability of the $m^{th}$ outcome is $p_{m}=|\psi_{m}|^2$, where $\psi_{m} = \langle m | \Psi\rangle$ is the coefficient of the wave function in the computational basis. Consider a case where the measurement has been repeated $N$ times, giving the following outcomes $[00, 01, 00, \hdots]$. The number of times the $m^{th}$ outcome occurs in the dataset is labeled with $n_m$ where trivially $\sum_{m}n_{m}=N$. The likelihood function defined in Eq.~(8) for our set of measurements of the two qubit wave function is re-written as
	\begin{align}\label{cost-fn-freq}
		L &= \frac{1}{N}\sum_{i=1}^N \log{\left|\langle m^{(i)} | \Psi\rangle\right|^2} = \frac{1}{N}\sum_{i=1}^N \log p_{m^{(i)}} \nonumber \\
		&= \frac{1}{N} (n_{00} \log p_{00} + n_{01} \log p_{01} + n_{10} \log p_{10} + n_{11} \log p_{11} ) \nonumber \\
		&= \frac{1}{N} \log \left( \prod_m\ p_m\ ^{n_m} \right) = \frac{1}{N} \log \mathcal L
	\end{align}
	where $m^{(i)}$ is the outcome of the $i^{th}$ measurement in the set of $N$ measurements. Using the technique of Lagrange multipliers, $L$ can be maximized under the constraint $\sum_m p_m = 1$ which stated below,
	\begin{equation}
		\frac{ \partial (\mathcal L - \alpha \sum_m p_m)}{\partial p_{m'}} = \frac{n_{m'}}{p_{m'}}\mathcal L - \alpha = 0 \qquad \forall\ m'.
	\end{equation}
	The factor ${\mathcal L}/{\alpha}$ is a constant that can be fixed by enforcing the condition $\sum_m p_m = 1$ giving 
	\begin{equation}
		p_m = \frac{n_m}{N}.
	\end{equation}
	Thus $L$ is maximized when $p_m = n_m/N$ implying that the maximum likelihood approach reduces to the frequentist method when measurements only along one direction are considered. 
	
	\subsection{Shadow Tomography}\label{shadowtom}
	The shadow tomography on the ensemble of states is performed similarly to the procedure described in \cite{o2016efficient,huang2020predicting,McGinley}. For every simulated measurement, the subsystem $B$ is measured in the $Z$ basis, whereas each qubit A is measured in  a random basis uniformly sampled from the set $\{X, Y, Z\}$. Each qubit in subsystem A is collapsed into some state $a_i$, where $i$ indexes the measured qubit, which depends both on the chosen basis and the measurement outcome. The datasets $\mathcal{D}(z_B)$ contains measurements of the state $|\Psi(z_B)\rangle$ in different basis. We label with $|\bm a^{(n)}\rangle$ the state that subsystem $A$ is collapsed into upon $n^{th}$ measurement and with $|a_i^{(n)}\rangle$ the collapsed state of qubit $i$. We consider total $M$ measurements and for each outcome $z_B$, the dataset $\mathcal{D}(z_B)$ contains many shots of subsystem $A$ measured in random local bases. More specifically, 
    \begin{equation}
        \mathcal{D}(z_B):= \{(s_n, b_n)\},
    \end{equation}
    where $s_n$ records the random Pauli basis selected on each qubit and $b_n$ denotes the measurement outcome of $\pm1$. 
    From a single element in $\mathcal{D}(z_B)$,  we build a single shadow $\hat{\rho}_A(z_B)$ via the inverse measurement map 
    \begin{equation}\label{eq: classical shadow}
        \hat{\rho}_A^{(n)}(z_B) = \bigotimes_i \left( 3 |a_i^{(n)}\rangle\langle a_i^{(n)}| - I\right),
    \end{equation}
    derived in \cite{huang2020predicting}.
    
     In general, $\rho_A^{\otimes k} (z_B)$ can be typically estimated using U-statistics of order $k$, which provides an unbiased approach to estimate a parameter, defined as
     	\begin{eqnarray}
     		U_k = \binom{M}{k}^{-1} \sum_{1 \leq i_1 < \cdots < i_k \leq M} 
     		h(\mathbf{X}_{i_1}, \mathbf{X}_{i_2}, \dots, \mathbf{X}_{i_k}), \nonumber\\
     	\end{eqnarray}
     	where $\mathbf{X}_{i_k}\in \{X_1, X_2, ..., X_n\}$ are samples from a distribution, and $h$ is a symmetric kernel function. Here, $X_n =  \hat{\rho}_A^{(n)}(z_B)$ and for simplicity, let us consider two cases: $k=1$ and $k=2$, for which the estimates of $\rho_A(z_B)$ and $\rho_A^{\otimes 2} (z_B)$ are given by
     	\begin{eqnarray}
     		{\rho}_A(z_B)&=&\frac{1}{M}\sum\limits_{n=1}^{M}\hat{\rho}_A^{(n)}(z_B), \label{U1} \\
     		{\rho}_A^{\otimes 2}(z_B)&=&\frac{1}{M(M-1)}\sum\limits_{n \neq m}^{M}\hat{\rho}_A^{(n)}(z_B) \otimes \hat{\rho}_A^{(m)}(z_B), \hspace{25pt} \label{U2}
     	\end{eqnarray}
     	which can be generalized to the $k^{th}$ moment. However, with an increase in $k$, the computational complexity scales as $M^k$. 
     	
     	     	    \begin{figure}[t]
     		\centering
     		\includegraphics[width=0.79\linewidth]{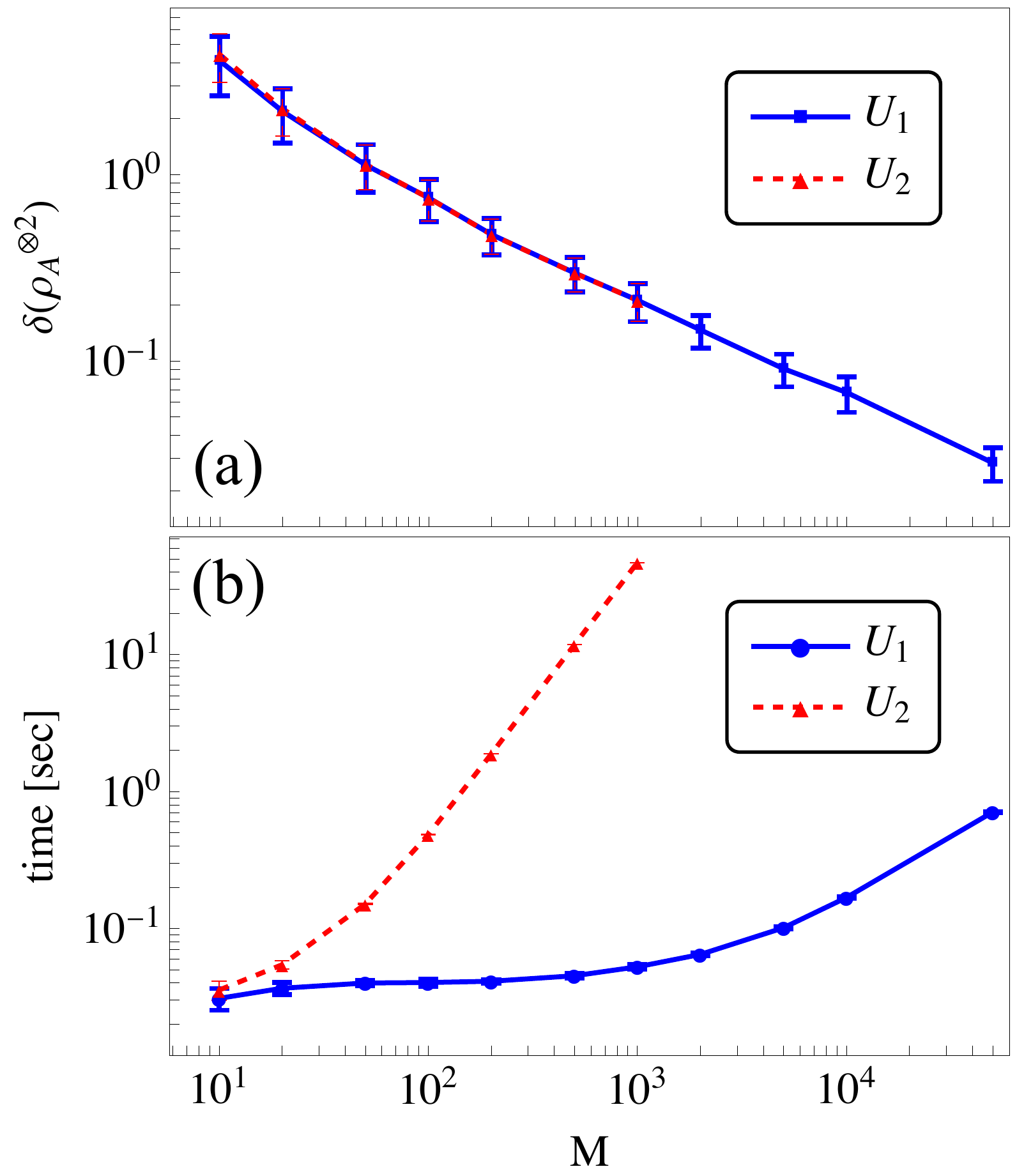}
     		\caption{ (a) Comparing trace distance from the exact density matrix from quantum evolution ($(\rho_A^{\rm exact})^{\otimes k}$) with estimate in Eq.~\ref{U2} ($U_2$ statistics) and Eq.~\ref{rho2_U1} ($U_1$ statistics)  shown with red-triangle and blue-circle, respectively. (b) Comparison of computation time of $U_1$ (blue-circle) and $U_2$(red-triangle) statistics for estimation of $\rho_A^{\otimes 2}$ with increase sample size. Note: Results for computation time are based on calculations performed on Intel core Ultra 7 265K.}
     		\label{fig:U1 vs U2}
     	\end{figure}
     	
     	Thus, in order to reduce the computational time, we consider a ``plug-in estimator" approach, where we consider copies of Eq.~\ref{U1} following $U_1$ instead of $U_2$ statistics, which provides us an approximate estimate $ \bar{\rho}_A^{\otimes 2}(z_B) = ({\rho}_A(z_B))^{\otimes k}$. To identify the regime where the approximation holds, we can derive a relation between the two estimators $ \bar{\rho}_A^{\otimes 2}(z_B)$ and   ${\rho}_A^{\otimes 2}(z_B)$. To begin with, we expand $ \bar{\rho}_A^{\otimes 2}(z_B)$ as follows
     	\begin{eqnarray}
     		 \bar{\rho}_A^{\otimes 2}(z_B) &=& ({\rho}_A(z_B))^{\otimes k},\\
     		 &=&\bigg(\frac{1}{M}\sum\limits_{n=1}^{M}\hat{\rho}_A^{(n)}(z_B)\bigg)\otimes\bigg(\frac{1}{M}\sum\limits_{m=1}^{M}\hat{\rho}_A^{(m)}(z_B)\bigg), \nonumber \\
     		 &=& \frac{1}{M^2}\sum\limits_{n,m=1}^{M}\hat{\rho}_A^{(n)}(z_B)\otimes\hat{\rho}_A^{(m)}(z_B), \nonumber \\
     		 &=& \frac{1}{M^2} \bigg(\sum\limits_{n\neq m}^{M}\hat{\rho}_A^{(n)}(z_B)\otimes\hat{\rho}_A^{(m)}(z_B) \nonumber\\
     		 & &+\sum\limits_{n=1}^{M}\hat{\rho}_A^{(n)}(z_B)\otimes\hat{\rho}_A^{(n)}(z_B)\bigg). \label{rho2_U1}
     	\end{eqnarray}
     	Using the expression of $\rho_A^{\otimes 2}$ from Eq.~\ref{U2}, we get
     		\begin{eqnarray}
     		\bar{\rho}_A^{\otimes 2}(z_B)
     		&=&\frac{M(M-1)}{M^2}  \rho_A^{\otimes 2} + \frac{1}{M^2}\sum\limits_{n=1}^{M}\hat{\rho}_A^{(n)}(z_B)\otimes\hat{\rho}_A^{(n)}(z_B), \nonumber \\
     		&\approx&  \rho_A^{\otimes 2}+\underbrace{ \frac{1}{M^2}\sum\limits_{n=1}^{M}\hat{\rho}_A^{(n)}(z_B)\otimes\hat{\rho}_A^{(n)}(z_B)}_{\rm Bias}. \label{U1_U2_diff}
     	\end{eqnarray}
     	Thus, from Eq.~\ref{U1_U2_diff}, the approximation is valid in the limit of sufficiently large measurements $M$.  To demonstrate this, Fig.~\ref{fig:U1 vs U2}~(a) compares the trace distance from the exact density matrix from quantum evolution ($(\rho_A^{\rm exact})^{\otimes k}$) with estimate in Eq.~\ref{U2} (based on $U_2$ statistics, shown as red triangles) and Eq.~\ref{rho2_U1} (based on $U_1$ statistics, shown as blue circles). The two curves become indistinguishable for sufficiently large $M$, indicating that it is sufficient to use $U_1$ statistics for estimating $\rho_A^{\otimes k}$.
     	
     	We further compare the computational run-time in Fig.~\ref{fig:U1 vs U2}~(b), which highlights the computational efficiency of the $U_1$ estimator relative to the unbiased $U_2$ estimator.  The added computational cost in $U_2$ arises from $M(M-1)$ combinations required for the tensor product in  Eq.~\ref{U2}, as compared to only $M$ combinations required in $U_1$ statistics in Eq.~\ref{U1} and \ref{rho2_U1}, as discussed earlier. Thus, for constructing a $k$-design, using $U_1$ statistics provide an $M^{k-1}$ advantage over the corresponding $U_k$ statistics.

%

\end{document}